\documentclass[nologo,url,11pt,a4paper]{ETHpaper}
\usepackage{verbatim}
\usepackage{graphicx}
\usepackage{subfigure}

\usepackage{amsfonts}
\usepackage{array}
\usepackage{amsthm}
\usepackage{amsmath}

\usepackage{appendix}

\newcommand{\mean}[1]{\left\langle #1 \right\rangle}
\renewcommand*{\=}{{\kern0.1em=\kern0.1em}}
\renewcommand*{\-}{{\kern0.1em-\kern0.1em}}
\newcommand*{\+}{{\kern0.1em+\kern0.1em}}

\begin{document}

\title{How big is too big? Critical Shocks for Systemic Failure Cascades}

\titlealternative{How big is too big? Critical Shocks for Systemic Failure Cascades}

\author{Claudio J. Tessone,  Antonios Garas, Beniamino Guerra, Frank
  Schweitzer\footnote{Corresponding author: \url{fschweitzer@ethz.ch}}
}

\authoralternative{C.~J. Tessone,  A. Garas, B. Guerra, F. Schweitzer}

\address{Chair of Systems Design, ETH Zurich, Weinbergstrasse 58, CH-8092
  Zurich, Switzerland}

 \reference{Submitted (2012). }

\www{\url{http://www.sg.ethz.ch}}

\makeframing
\maketitle

\begin{center}
 \date{\today}
\end{center}

\begin{abstract}
  External or internal shocks may lead to the collapse of a system consisting of many agents.  If the shock hits only one agent initially and causes it to fail, this can induce a cascade of failures among neighboring agents. Several critical constellations determine whether this cascade remains finite or reaches the size of the system, i.e. leads to systemic risk. We investigate the critical parameters for such cascades in a simple model, where agents are characterized by an individual threshold $\theta_{i}$ determining their capacity to handle a load $\alpha\theta_{i}$ with $1-\alpha$ being their safety margin. If agents fail, they redistribute their load equally to $K$ neighboring agents in a regular network. For three different threshold distributions $P(\theta)$, we derive analytical results for the size of the cascade, $X(t)$, which is regarded as a measure of systemic risk, and the time when it stops. We focus on two different regimes, (i) \emph{EEE}, an external extreme event where the size of the shock is of the order of the total capacity of the network, and (ii) \emph{RIE}, a random internal event where the size of the shock is of the order of the capacity of an agent. We find that even for large extreme events that exceed the capacity of the network finite cascades are still possible, if a power-law threshold distribution is assumed. On the other hand, even small random fluctuations may lead to full cascades if critical conditions are met. Most importantly, we demonstrate that the size of the ``big'' shock is not the problem, as the systemic risk only varies slightly for changes of 10 to 50 percent of the external shock. Systemic risk depends much more on ingredients such as the network topology, the safety margin and the threshold distribution, which gives hints on how to reduce systemic risk.
\end{abstract}

\section{Introduction}

Current research on systemic risk can be roughly divided into two
different strands each one having its own focus: (i) the probability
of extreme events which can cause a breakdown of the system, (ii) the
mechanisms which can amplify the failure of a few system elements, to
cause a failure cascade of the size of the system. The former line of research
assumes that systemic risk is caused by \emph{external} events,
e.g. big earthquakes, tsunamis, or meteor impacts. Thus, in addition
to the likelihood of extreme events, another interesting question regards
the response of the system to such perturbations, i.e. its capability
to absorb shocks of a given size.  The latter research area, on the other
hand, sees systemic failure as an \emph{endogenous} feature that
basically emerges from the non-linear interaction of the constituents, i.e. how they redistribute, and possibly amplify, load internally.

In both approaches, the likelihood of a systemic breakdown can only be determined by considering the internal dynamics of system elements, denoted as \emph{agents} in this paper,
such as their capacity to resist shocks, their time-bound
interaction with neighbors, their dependence on macroscopic feedback
mechanisms, such as coupling to the macroscopic state of the system. Only in rare cases the
dynamics of systemic risk can be reduced to  mere topological
aspects, such as the diversity in the number of neighbors, the role of
hubs, \textit{etc.}.
In this paper, we combine the two research questions mentioned above:
on the one hand, we are interested in the critical size of an external
shock that may lead to collapse of the system. At the same time, we
address that such critical values depend on the safety margins of the
system elements, and the details of their interaction when
redistributing load internally.
We also investigate how these \textit{cascading dynamics} are affected
by the structural features of the network (level of connectivity,
topological heterogeneities)and by individual properties of the agents, such as the probability
distributions of the \textit{failure thresholds}.  Such insights can
directly benefit a robust system design by means of individualization
of agents (i.e. designing agents with optimal heterogeneity).

Given the importance of such problems for social, economical and
technological systems, the topic is already discussed in a wide range
of scientific literature.  Some modeling framework were recently
proposed \cite{Lorenz2009b, Watts2002SimpleModelof,
  dodds2004universal}.  The \textit{complex network} approach was also
used to describe cascading processes in power grids and in Internet
services \cite{motter2002cascade, crucitti2004model}, and was also
applied to data storage services \cite{rossi2009}. Importantly,
similar agent-based approaches were developed to model avalanche
defaults among financial institutions
\cite{Battiston2009a,gai2010contagion}.

Our paper is organized as follows: in the next section we introduce
the agent-based model studied, determining e.g. agent's {\em
  fragility} and agents' interaction by means of a load redistribution
mechanism. This allows us to define a measure for {\em systemic risk}
on the macroscopic level.  In Section \ref{sec:risk} we develop an
analytical framework that allows us to unveil the dynamics of systemic
risk based on cascading processes.  In Section \ref{sec:het}, we
discuss the critical conditions for systemic risk to emerge.  Later,
in the Section \ref{sec:sysnodes}, we study when agents can be
considered as {\em systemic} by their importance. The paper finishes
with some conclusions in Section \ref{sec:concl}, which also allow for
a generalized picture of how to prevent systemic risk.

\section{Analytical approach to systemic risk} \label{sec:analytical}

\subsection{Description of the model}

\paragraph{Net fragility}

In a recent paper \cite{Lorenz2009b}, a framework
to model systemic risk by means of an agent-based approach was developed.
In this framework, each agent $r$ is characterized by three individual
variables: a discrete variable $s_{r}(t)\in\{0,1\}$, which describes
its state at a discrete time $t$, i.e. $s_{r}(t)=0$ for an operating
state and $s_{r}(t)=1$ for a failed state, and two continuous
variables, the threshold $\theta_{r}$ and load $\phi_{r}$. The
threshold is assumed to describe the individual `\emph{capacity}' of an
agent: it defines how much load an agent can carry before it fails.
On the other hand, the variable $\phi_r$ describes the {\em load} which is exerted on an agent.

We note that while the load can change in time e.g. through systemic feedback, it
further depends on the state of other agents $\underline{s}$ and on
the network of interactions, described by the adjacency matrix $A$.
Written this way, the load also depends on how it is exchanged between agents.
A special case will be discussed below.
We define that agent $r$ fails if its \textit{net fragility} $z_r(t)$,
 \begin{equation}
  z_{r}(t)=\phi_{r}(t,A,\underline{s})-\theta_{r},
  \label{eq:netfragility}
\end{equation}
is equal or larger than zero.
I.e, in a deterministic model, the dynamics of an agent is given by
\begin{equation}
  s_{r}(t+1)=\Theta \big( z_{r}(t) \big),
  \label{eq:detdyn}
\end{equation}
where $\Theta( \cdot )$ is the Heaviside function.
Certainly, the dynamics only depends on the net fragility, i.e.~on the
relative distance between load and threshold.
Nevertheless, a distinction between these two individual variables is very useful, as
it allows us to conceptually distinguish between internal and external influences on the failure.

\paragraph{Systemic risk}

We now define the important measure of systemic risk. We define it as the fraction of
failed agents at any point in time. For a system composed of $N$ agents, it reads
\begin{equation}
  X(t)=\frac{1}{N}\sum_{r}s_{r}(t) =\int_{0}^{\infty} p_{z(t)} \big( z \big) dz;
\end{equation}
$p_{z(t)} \big( z \big)$ represents the density of agents with a net fragility $z$ at time $t$; the integral runs over the agents whose net fragility is positive.
Failures in a subset of agents will result in cascading processes over the
network of interaction, which results in changes of the fragility of
other agents in the course of time.
This can be expressed by the recursive dynamics $p_{z(t+1)}=\mathcal{F}(p_{z(t)})$, where
$\mathcal{F}$ is some function that describes how the load of failing
agents is redistributed depending on the interaction mechanisms.
With this, by specifying the initial condition $p_{z(0)}$, it is possible to compute
$X(t)$ for a deterministic dynamics.

In Ref.~\cite{Lorenz2009b}, $X(t)$ was calculated by making
suitable assumptions about the distribution of the net fragility,
$p_{z}(t)$, the initial conditions $p_{z(0)}$, and for the particular case of a fully connected network -- i.e.~each agent interacts with
everyone else. Specifically, an initial condition $p_{z(0)}\sim
\mathcal{N(\bar \theta,\sigma)}$, was used; i.e. the initial fragility of agents is
normally distributed with a mean $\bar \theta$ and standard deviation $\sigma$.
This implies that the initial fraction of failed agents at time $t=0$ is
given by $X(0)=\Phi_{\bar \theta,\sigma}(0)$, where $\Phi_{\bar \theta,\sigma}(0)$
denotes the cumulative function of the normal distribution. I.e. it gives the (normalized)
number of agents with an initial net fragility (defined in Eq.~\eqref{eq:netfragility}), equal or larger than zero.
The authors calculated the size of cascades
measured by the final fraction of failed agents for different
interaction mechanisms.
Remarkably, it was found that systemic risk
depends on the variance $\sigma$ of the distribution $p_{z}(0)$ in a non-monotonous
manner.
This means, systemic risk can decrease if the agents become
more heterogeneous, i.e. if their individual threshold becomes more
different. On the other hand, for homogeneous agents characterized by
the same threshold, a first-order phase transition was found between no
systemic risk and complete failure.

\paragraph{Initial fragility}

We use these previous findings as a reference point, but we will extend our model
in different ways.
First of all, instead of a normal distribution for the
initial net fragility $z_{r}(0)=\phi_{r}(0)-\theta_r$, we assume a fixed
relation between initial fragility $\phi_{r}(0)$ and threshold
$\theta_{r}$:
\begin{equation}
 \phi_{r}(0)=\alpha \theta_{r} \quad (r=1,\ldots,N; \; r\neq i).
\label{P_c_+uniform}
\end{equation}
The parameter $\alpha$ is a constant, equal for all agents.  Only for
one agent $i$, instead of the fixed relation (\ref{P_c_+uniform}), we
define $\phi_{i}(0)=\phi_{\star}\geq  \theta_{i}$.  Thus, we
consider that initially only one agent $i$, is at a critical
condition, whereas with $\alpha<1$ all other agents are initially
capable of handling the load assigned to them.  I.e. different from the
distribution of initial loads in \cite{Lorenz2009b}, we do not have an
initial failure cascade.  Instead,
the initial condition for the systemic risk is simply $X(0)=1/N$.

The value $1-\alpha$ can then be regarded as the  {\em agent available
  capacity} (or safety margin) before they fail if their load is
increased. A fixed relation between fragility and threshold was first
used in \cite{motter2002cascade} to describe cascading processes in
power grids and Internet (see also \cite{crucitti2004model,
  Motter2004CascadeControland, CRUCITTI_cascades_power_grid}). It
basically reflects the situation of many socio-technical systems in
which the capacity of agents is usually \textit{ad hoc} designed to
handle the load, because limited by cost, under normal conditions.
We will later vary the safety margin $1-\alpha$ to determine how the
severity of cascading failures will depend on it.

\paragraph{Threshold distribution}

With these considerations, only the threshold distribution $P(\theta)$
remains to be specified to complete the initial conditions.
It is worth remarking that the capacities of the agents --in contrast with other studies found in the Literature so far-- are decoupled from the topological artifacts of the network connecting them.
Here we
will use three different assumptions
for both analytical calculations and computer simulations:
\begin{itemize}
\item[(a)] a delta distribution $P(\theta)=\delta(\theta-\bar{\theta})$, where $\delta(\cdot)$ is the Dirac delta function,  i.e. all agents have the same threshold $\bar{\theta}$,
\item[(b)] a uniform distribution $P(\theta)=U(\bar \theta-\sigma,\bar
  \theta+\sigma)$ with the mean $\bar \theta$ and the range $\sigma$,
  i.e. all agents have different, but comparable thresholds in the
  interval $[\bar \theta-\sigma,\bar \theta+\sigma]$.
  For all further calculations we define $\theta_{\mathrm{min}}=\bar{\theta}-\sigma$.
\item[(c)] a power-law distribution
\begin{equation}
P(\theta) = \frac{\gamma -1}{\theta_{\mathrm{min}}^{1-\gamma}}\, \theta^{-\gamma}
\label{eq:power}
\end{equation}
i.e. agents have thresholds that can differ by orders of
magnitudes.
As the normalization depends on the value $\theta_{\mathrm{min}}$, we
assign its \emph{numerical value} for our further calculations to be
comparable with the minimum value of the uniform distribution,
$\theta_{\mathrm{min}}=\bar{\theta}-\sigma$.
\end{itemize}

\paragraph{Agent interaction}

In order to describe the agent interaction, we use the network
approach in which agents are represented by nodes and interactions by
links between agents. I.e., the network topology specifies which other
agents a particular one interacts with. This can be statistically
described by the degree distribution $P(k)$ for which we will use in
this paper only $P(k)=\delta_{k_{r},K}$, i.e. a regular network, in
which each agent interacts with $K$ other agents.
The fully connected network is a special case with $K=N-1$.

\begin{figure}
  \centering
\includegraphics[width=0.65\textwidth,angle=0]{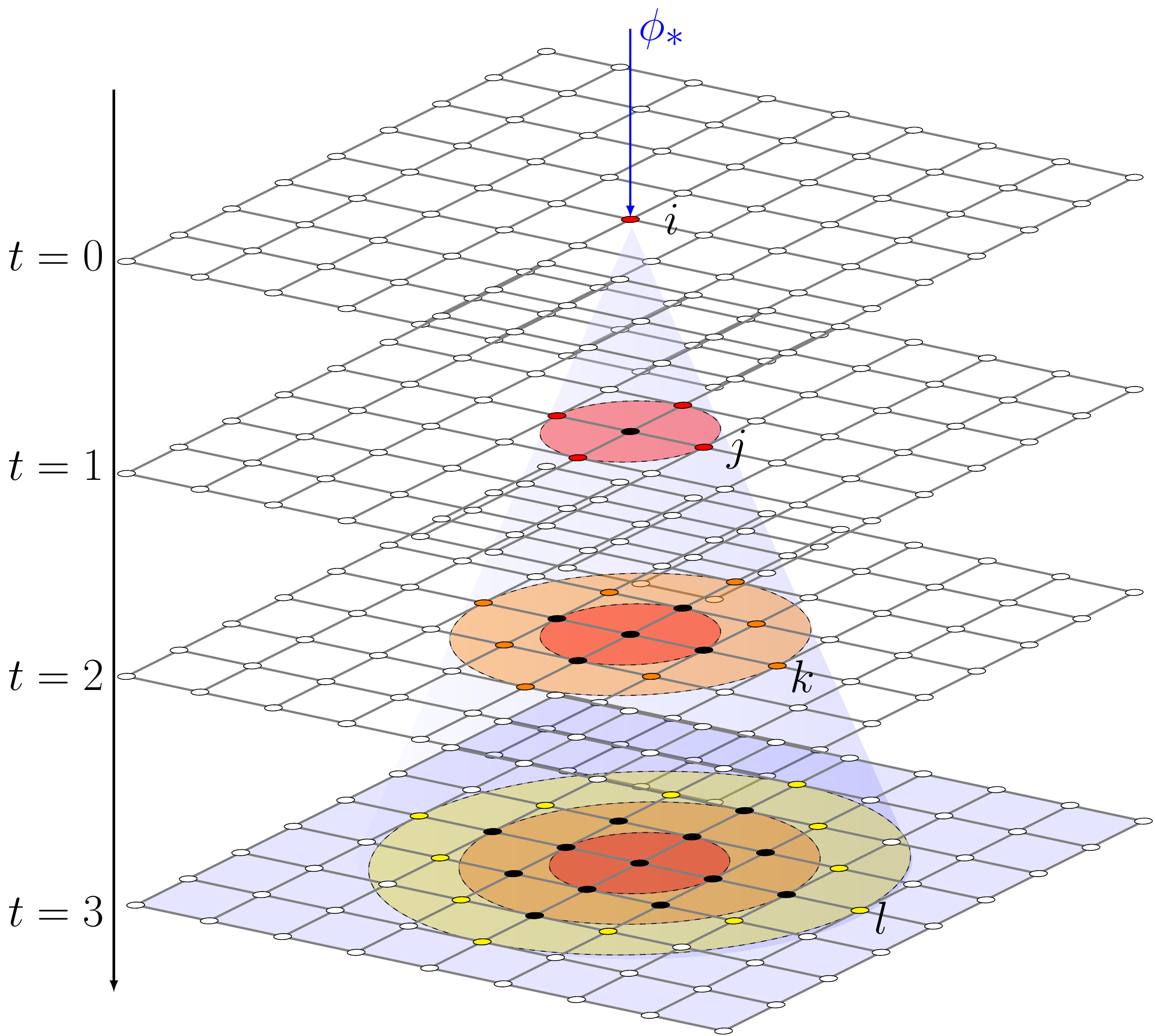}
\caption{ Regular lattice with $K=4$, where agent $i$ is hit
    initially ($t=0$) by an external shock of size $\phi_{\star}$. If $i$
    fails, it distributes this load to its $K$ nearest neighbors in
    the next time step ($t = 1$). If they fail, they distribute their load to their $K$
    nearest neighbors in the next time step ($t = 2$) , which are the $2K$ second
    nearest neighbors of $i$, \textit{etc.}. }
  \label{fig:lattice}
\end{figure}

Secondly, we have to specify how agents interact through these
links. Here, we assume a load redistribution mechanism in which the
initially failing agent $i$ shares its load $\phi_{i}(0)\equiv
\phi_{\star}$ equally among its $K$ neighboring agents (labeled $j$),
see Fig. \ref{fig:lattice}. That means for each of these agents, their
own load $\phi_{j}(t)$ increases in the next time step $(t+1)$ by an
amount of $\phi_{\star}/{K}$. If this addition leads to a positive net
fragility $z_{j}(t)=\phi_{j}(t)-\theta_{j}$, agents $j$ fail as well
and redistribute their load $\phi_{j}(t+1)$ equally to their $K$
neighboring agents, and so on.

\paragraph{Cascade sizes}

This way, failure cascades can occur in the course of time, and we are
interested in their relative size and the probability distribution of
their occurrence, $P(X(t))$. For this calculation, which will be done
in Sect. \ref{sec:het}, we define the fraction $f(t)$ of failing
agents at each time step $t$ and the number $F(t)$ of failing agents
during the same time interval as:
\begin{equation}
  \label{eq:ff}
  f(t)=\frac{1}{K(t)}\sum_{r=1}^{N} s_{r}(t)-s_{r}(t-1)\;;\quad F(t)=K(t)\,f(t)
\;;\quad f(0)=1.
\end{equation}
$K(t)$ gives the number of agents that are hit by the cascade at
time $t$, i.e.~they are located in the $t$--th neighborhood of agent $i$
which failed initially. Hence, with the model of
Fig. \ref{fig:lattice} in mind, $K(t)$ is the number of agents that
can {\em potentially} fail during time step $t$.  Dependent on the
topology of the regular network, there are two limiting cases to
express how $K$ grows in time. $K(t) \propto K\, t$ holds in regular
networks, where the interface grows linearly with distance.  On the
other hand, for Bethe lattices, tree-like structures and random
topologies in which loops are neglected \cite{Watts2002SimpleModelof},
the number of nodes at distance $t$ is $K(t) = K^t$.

Using the definition (\ref{eq:ff}), we can calculate the size of the
cascade at time $t$, which is equal to the systemic risk $X(t)$ as:
\begin{equation}
  \label{eq:xt}
  X(t)=\frac{1}{N}\sum_{\tau=0}^{t} F(\tau)= \frac{1}{N}
  \sum_{\tau=1}^{t} K(\tau)\, f(\tau).
\end{equation}
In general, Eq.~(\ref{eq:xt}) cannot be solved analytically.
However, in the following sections we will derive analytical
expressions for $f(t)$, assuming different distributions of agents'
thresholds.

\paragraph{Finite versus infinite cascades}

According to the definitions above, a total failure occurs if $X(t)=1$. In a finite system, this will happen at a finite time $t^{\prime}$, while in an infinite system this final state is reached only asymptotically, $t^{\prime}\to \infty$. However, in a finite system  a cascade can stop even for $X(t)<1$ if the potential number of failing agents reaches the system size at a given time $t^{\circ}$,
\begin{equation}
\sum_{t=0}^{t^\circ} K(t) \geq N.
\label{tcirc}
\end{equation}
Yet, there is a third case to be considered, namely that the cascade
stops at a finite time $t^{\star}$, even if $X(t)<1$ and
$t<t^{\circ}$, simply because the redistribution of loads to the
nearest neighbors does not cause further failure. This is expressed by
the condition $f(t^{\star})=0$.

We will refer to an ``infinite'' cascade if $X(t^{\prime})=1$, which
means every agent in the system has failed. On the other hand, a
``finite'' cascade occurs either if it stops at time
$t^{\star}<t^{\circ}$, or if the redistribution of load has reached
the system size, Eq.~(\ref{tcirc}), without causing all agents to
fail. Consequently, finite cascades stop at time $t^{\prime} =
\min(t^\circ,t^{\star})$. We note that, according to our definition,
Eq.~(\ref{eq:xt}), systemic risk refers to finite cascades as well,
not just to $X\to 1$. Precisely, we are interested in the distribution
$P(X(t^{\prime}))$, i.e. the density of failed agents at the time by which cascades end regardless of the cause for that.

\begin{figure}
\centering
\includegraphics[width=0.6\textwidth,angle=0]{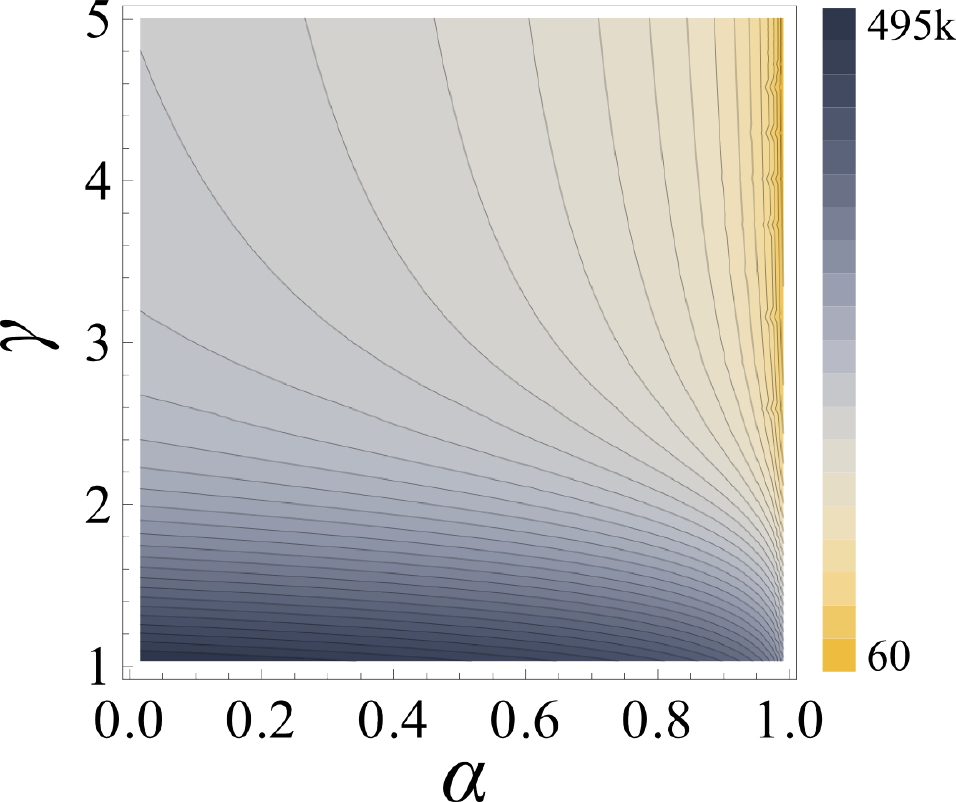}
\caption{Contour plot showing the network capacity $Q$ as a function of $\alpha$ and $\gamma$, for a network with $N=1000$ nodes having power-law threshold distribution with a minimum threshold value $\theta_{\mathrm{min}}=0.5$.
}
\label{fig:Q-plaw}
\end{figure}

\paragraph{Network capacity}
In order to put the size of the initial shock $\phi_{\star}$ into perspective, we refer to the total capacity of the network to absorb shocks, which depends on the safety margin $(1-\alpha)$, the total number of nodes, and the threshold distribution $P(\theta)$.
Thus, the capacity $Q$ that the system could {\em a priori} absorb during the cascade is simply given by
\begin{equation}
Q = N (1-\alpha) \int d\theta \, \theta \, P(\theta).
\label{eq:q}
\end{equation}
If the threshold distribution has a defined mean value, $\bar \theta$,
this expression reduces to \mbox{$Q = N (1-\alpha)\, \bar \theta$}.
On the other hand, for a normalized power-law distribution with a
minimum threshold value $\theta_{\mathrm{min}}$, the mean value is
only defined for $\gamma > 2$. For $\gamma \leq 2$, a simple argument
\cite{Tessone2010} shows that for a finite system the expected value
can still be computed. The result is
\begin{equation}
\frac{Q}{N(1-\alpha)} = \left\{
\begin{array}{lll}
 \theta_{\text{min}}  & \left[\frac{\displaystyle \gamma-1}{\displaystyle \gamma-2}+  \frac{\displaystyle N^{2-\gamma}}{\displaystyle 2-\gamma} \right] & \text{if} \, \gamma \leq 2  \\
  \theta_{\text{min}}  &  \left[\frac{\displaystyle \gamma-1}{\displaystyle \gamma-2} \right] & \text{if}\, \gamma >2
\end{array}
\right. .
\label{qcases}
\end{equation}
It is worth noticing that for the delta threshold distribution, the
uniform (with $\bar\theta \sim \sigma$) and the power-law distribution with $\gamma > 2$, the network
capacity $Q$ is of the same order of magnitude.
In Fig.~\ref{fig:Q-plaw} we show the network capacity $Q$ for the power-law distribution as a function of $\gamma$ and $\alpha$.
Precisely, it gets the same \emph{numerical value} in all three cases cases,
$Q=Q_{u}$, if $\gamma=1.5$ and $\bar \theta = 2
\theta_{\mathrm{min}}$, as used for the numerical calculations.
However, for the power law distribution with $\gamma\leq 2$, the
network capacity becomes much larger than in the three other cases
because of the additional dependence of the number of agents,
$N^{2-\gamma}$. Choosing $\gamma=1.5$ for the numerical calculations
later implies that, compared to the uniform case, we have $Q\propto
\sqrt{N}Q_{u}$.

In this paper, depending on the magnitude of the initial shock, we distinguish between two different regimes:

(i) \textbf{EEE} -- the \emph{extreme exogenous event} resulting in a very
large $\phi_{\star}$ which is of the order of $Q$, i.e. much larger
than the capacity of the initially failing agent (or the average
capacity $\bar{\theta}$ of agents): $\phi_{\star}\sim Q \gg
\bar\theta$. In this case, there is no surprise that agents involved
in the redistribution of load will fail, at least in an early
phase. Hence, we are mostly interested in the conditions under which
cascades may stop before they have reached the size of the system.

(ii) \textbf{RIE} -- the \emph{random internal event}, which assumes that
initially one randomly chosen agent $i$ faces a load $\phi_{\star}$
that is slightly larger than its own capacity $\theta_{i}$, drawn from
the distribution $P(\theta)$, i.e. $\phi_{\star}\sim \bar\theta \ll
Q$. This is likely to happen by a random fluctuation of the load
$\phi_{i}$ that exceeds the threshold, rather than a big impact on the
system. In this case, we are interested in the conditions under which
cascades occur at all.

\subsection{Conditions for failing nearest neighbors}
\label{sec:condition}
We assume that, at $t=0$, a single randomly chosen agent $i\in 1
\ldots N$ fails because of an initial shock, i.e. $\phi_{i}(0) \equiv
\phi_{\star} \geq \theta_{i} $. According to the redistribution mechanism
described above, this failure will increase the fragility of the
nearest neighbors of $i$, labeled $j \in nn(i)$ (see
Fig.~\ref{fig:lattice}). Agent $j$ can fail if its net fragility
becomes positive, i.e.:
\begin{equation}
\phi_{j}(1)= \phi_{j}(0) + \frac{\phi_{\star}}{K}\geq \theta_j,
\label{fail}
\end{equation}
which together with Eq.~(\ref{P_c_+uniform}) leads to the critical
condition for the failure of agent $j$,
\begin{equation}
  \theta_j \leq \theta_{(1)}^{c}(\phi_{\star})= \frac{\phi_{\star}}{K(1-\alpha)}.
  \label{theta_c}
\end{equation}
Here $\theta_{(1)}^{c}(\phi_{\star})$ defines the critical threshold for
the first-order neighborhood of agent $i$, or the critical threshold
  at time $t=1$, respectively. Agents with a threshold between
  $\theta_{\mathrm{min}}$ and $\theta_{(1)}^{c}$ will fail, hence the
  fraction of failing agents at time $t=1$ reads:
\begin{equation}
f(1)=\int_{\theta_{\mathrm{min}}}^{\theta^{c}_{(1)}(\phi_{\star})} d\theta P(\theta).
\label{eq:f_1_general}
\end{equation}
This fraction depends on the threshold distribution $P(\theta)$, so
explicit calculations will be given in the next Section.
At the moment we just assume that at least one agent $j$ has failed,
i.e. there will be a cascade to the next neighborhood
(cf. Fig. \ref{fig:lattice}).

Let us denote failing agents in the first step by $j^{\star}\in
nn(i)$. Their load $\phi_j^{*}(t=1)>\theta_j$ will be redistributed to
their nearest neighbors labeled $k \in nn(j^{\star})$.  Following the
reasoning used for Eq.~\eqref{fail}, we obtain for the load of agents $k$
at time $t=2$:
\begin{equation}
\phi_k(2)=\phi_k(1)+\sum_{j^{\star} \in nn(k)}\frac{\phi_{j^{\star}}(1)}{K}=
\alpha \theta_k +\frac 1 K \sum_{j^{\star} \in nn(k)} \left( \phi_{j^{\star}}(0)+\frac{\phi_{\star}}{K} \right).
\label{eq:fail_2_a}
\end{equation}
The summation is performed over the whole set of failed agents $j^{*}$
that belong to the neighborhood of $k$, their load being
$\phi_{j^{\star}}(0)=\alpha\theta_{j^{*}}$.

The exact amount of agents $j^{\star} \in nn(k)$ depend on the topological
properties of the network considered. For example, in square and
hexagonal lattices, some second nearest neighbors of $i$, i.e. agents
$k$, have more than one link to agents $j$ in the nearest
neighborhood. E.g. for the hexagonal lattice, half of the agents at
level $k$ have two links to agents $j$, whereas the other half has
only one. In general, for regular lattices, this number will be
between one or two.  In this paper, we will restrict our analysis to
the case of a single failing node $j^{\star}$ in the neighborhood of $k$
which is the case for Bethe lattices or sparse random regular networks
\cite{construction}.  The theory can be extended for other regular
geometries in a straight-forward manner.  With this considerations in
mind, Eq.~\eqref{eq:fail_2_a} becomes
\begin{equation}
\phi_k(2) =\alpha \theta_k + \frac 1 K \left( \alpha \theta_{j^{\star}} + \frac{\phi_{\star}}{K} \right),
\label{eq:fail_2}
\end{equation}
From Eq.\eqref{eq:fail_2} we obtain the critical condition for the
failure of agent $k$ if its net fragility becomes positive:
\begin{equation}
    \theta_k \leq \theta_{(2)}^c = \frac{\theta_{j^{\star}}+{\phi_{\star}}/{K}}{K(1-\alpha)}.
\label{eq:theta_c_2}
\end{equation}
This expression for the critical threshold at $t=2$, i.e. in the
second-order neighborhood of $i$, is comprised of two redistribution
processes. On the one hand those from agents $j^{\star}$ failing at $t=1$ and, on the other, the
redistribution of the initial load $\phi_{\star}$ from agent $i$ failing at
time $t=0$. The fraction of failed agents at time $t=2$ ($k \in
nn(j^{\star})$) is then given by
\begin{equation}
\label{eq:f_t_2}
f(2)=\int_{\theta_{\min}}^{\theta^{c}_{(1)}(\phi_{\star})}d\theta_{(1)} P(\theta_{(1)}) \int_{\theta_{\min}}^{\theta^{c}_{(2)}(\phi_{(1)})}P(\theta_{(2)})d\theta_{(2)},
\end{equation}
where $\theta^{c}_{(2)}(\phi_{(1)})$ indicates that the critical
threshold at $t=2$ depends on the load of failing agents $j^{*}$ at
time $t=1$, which does not need to be equal for every $j^{\star}$, but
depends on the topology.

Using the same reasoning for the different time steps of the cascade,
we obtain a general expression for the fraction of agents failing
during time step $t$ (which are the neighbors of agents failing at
$t-1$):
\begin{equation}
\label{f_t_general}
f(t)= \int_{\theta_{min}}^{\theta^{c}_{(1)}(\phi_{\star})} d\theta^{(1)}
P(\theta^{(1)}) \cdots \int_{\theta_{min}}^{\theta^{(c)}_{(t-1)}(\phi_{(t-2)})}d\theta^{(t-1)}
P(\theta^{(t-1)})
\int_{\theta_{min}}^{\theta^{c}_{(t)}(\phi_{(t-1)})} d\theta^{(t)} P(\theta^{(t)}),
\end{equation}
The critical threshold $\theta^c(\phi_{(t)})$ at time $t$ depends on the
load $\phi_{(t-1)}$ as follows:
\begin{equation}
    \theta_{(t)} \leq \theta^{c}_{(t)} = \frac{\phi_{(t-1)}}{K(1-\alpha)}.
\label{eq:theta_c_t}
\end{equation}
This is a recursive equation, i.e. $f(t)$ depends on the load
redistributed by all the agents that failed along the path connecting
the initially failing agent $i$ with agents failing at time $t$.
However, Eq.~\eqref{f_t_general} cannot be computed in general, thus,
in the following sections, we will study some cases in which this
equation can be reduced and solved.

\subsection{Threshold approximation  in the RIE and EEE regimes}
\label{sec:saf}

Inequality (\ref{eq:theta_c_t}) is an important result to understand
the propagation of cascades, which holds for both regimes introduced
above, the \textbf{EEE} regime, where the external shock dominates the
dynamics, and the \textbf{RIE} regime, where small random events inside the
first failing agents may trigger the cascade. For both, we are able to
derive some general results even before specifying the threshold
distribution $P(\theta)$.

In the \textbf{RIE} case, $\phi_{\star}\sim \bar\theta \ll Q$ the
redistribution of loads, that means the network effect, plays the most
important role.
Note that pairs of agents connected through an edge have in general different capacities. If one of them fails, its neighbor is exposed to failure during the following time step. However, whether it fails or not will depend on: (a) the load redistributed from the failing agent; (b) its own capacity.
Let us assume that if a agent with capacity
$\theta_{(t-1)}$ fails, the total load induced on its neighbors is
$\phi_{(t)} = \alpha \theta_{(t)} + { \theta_{(t-1)}}/{ K }$. This
assumption neglects the contribution of agents that failed before the
time step $t-1$, which are terms of order $K^{-\tau}$, with $\tau\geq
2$. This implies that $\theta_{(t-1)}$ is the load distributed by the
 agent, i.e.~it is exactly its capacity and not more. Thus, the largest capacity of the failing agent at time $t$, $\theta^c_{(t)}$, depends on the capacity of the agent that failed on
the previous time step $t-1$, i.e.\begin{equation}
  \theta^c_{(t)}(\theta_{(t-1)}) = \frac{ \theta_{(t-1)}}{ K (1-\alpha)}.
  \label{eq:rec:red1}
\end{equation}
which is a lower bound for the failure condition,
Eq. (\ref{eq:theta_c_t}). This means that among all neighbors of the agents at layer
$(t-1)$, those with a capacity lower than
$\theta^c_{(t)}(\theta_{(t-1)})$ will fail.

With the assumption \eqref{eq:rec:red1}, the fraction of failing agents at time $t$ can be decomposed in
terms of failure of two consecutive agents in a pair-wise approximation as follows:
\begin{equation}
 f(t) = \int_{\bar \theta - \sigma}^{\theta^c_{(t-1)}} d
 \theta_{(t-1)} \, P(\theta_{(t-1)}) \int_{\bar \theta - \sigma}^{\theta^c_{(t)}(\theta_{(t-1)})} d \theta_{(t)} \, P(\theta_{(t)}). \label{eq:rec:aux}
\end{equation}
This approach differs from the previous mean-field approximation, in the following. Now, the net effect of the load redistributed by a failing agent is taken into account to determine the fraction of its neighbors that will fail in the next time step. Thus, this approximation entails information about the heterogeneity at the edge level.
On the other hand, even in the \textbf{EEE} regime, $\phi_{\star}\sim Q \gg
\bar\theta$, the role of the network still cannot be neglected.
We are interested in the limit satisfying: (i) the load redistributed by the previously failed nodes cannot be totally neglected --i.e.~$\alpha\sim 1$--; (ii) the main contribution to the load $\theta_{(t)}$ comes from  $\phi_{\star}$ (i.e.~from the initial load).
We assume
that at any point in time, there exists a critical threshold
 $\theta_{(t)}^c$ above which agents do not fail.
 In this case
Eq.~\eqref{eq:theta_c_t}, becomes simply $\theta^c_{(t)} =
\theta^c_{(t-1)}/[K(1-\alpha)] $. Then, the load of agents at a
distance $t$ from agent $i$ is simply given by
\begin{equation}
\theta_{(t)}= \frac{{\phi}_{(t-1)}}{K(1-\alpha)} =\frac{{\phi_{\star}}}{[K (1-\alpha)]^{t}}.
\label{eq:theta-c-n}
\end{equation}
which results in the critical threshold for the \textbf{EEE} regime:
\begin{equation}
    \theta_{(t)} \leq \theta^{c}_{(t)}(\phi_{\star})
= \frac{\phi_{\star}}{[K(1-\alpha)]^{t}}.
\label{eq:theta_c_t-EEE}
\end{equation}

\begin{figure}
\centering
\includegraphics[width=0.6\textwidth,angle=0]{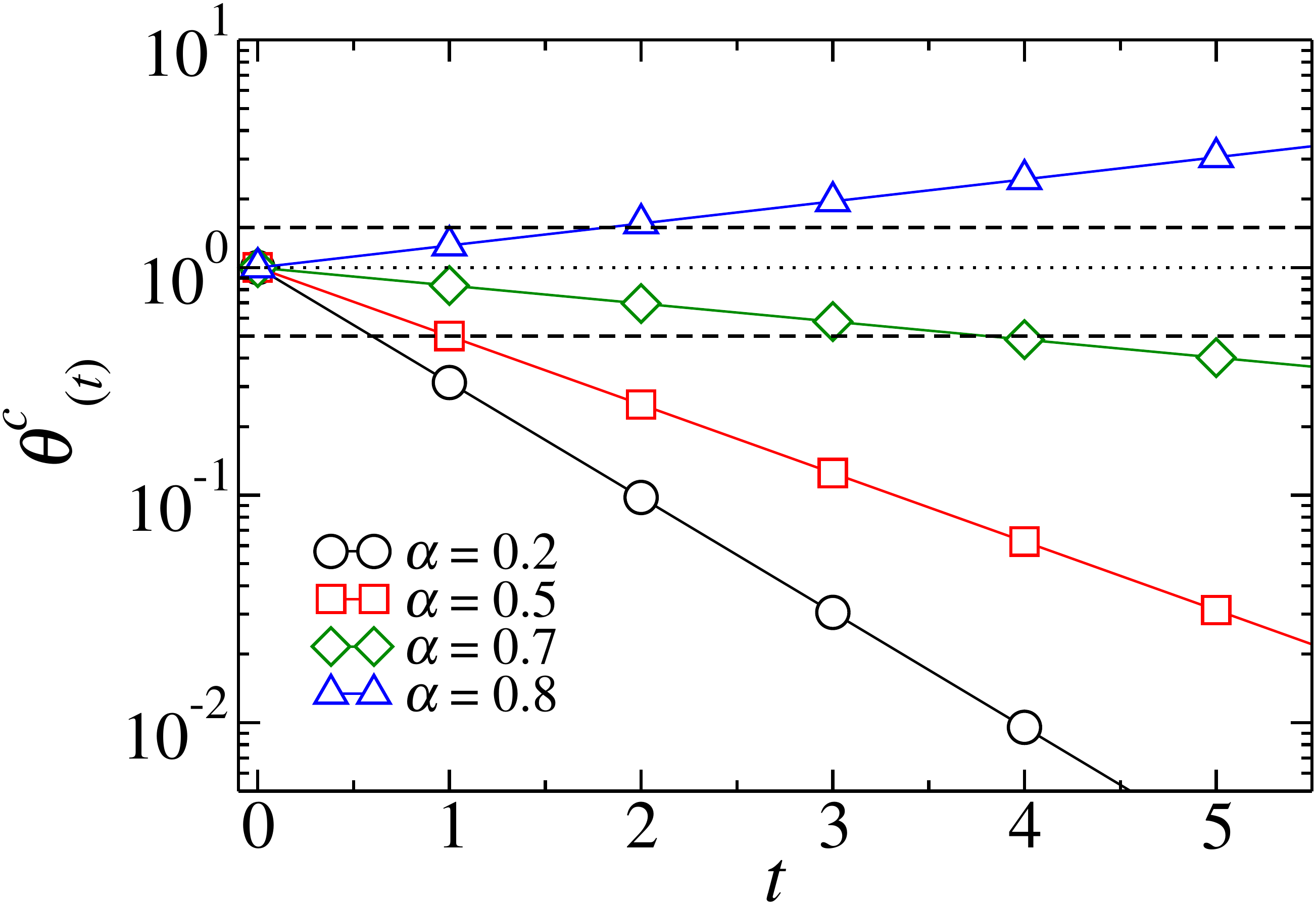}
\caption{
Evolution of the critical threshold $\theta^{c}_{(t)}$ (note the $\log$ scale) versus
  the time step $t$ for the case of a uniform distribution with parameters
  $\bar \theta=1$ (dotted line), and $\sigma=0.5$ (dashed lines).  The
  value of $K$ is set to four, while we used different values for the
  parameter $\alpha$.
  }
\label{fig:theta-c-n}
\end{figure}

This gives the critical condition for the failing threshold of an
agent that is hit by the cascade at time $t$ (i.e. it belongs to the
$t$-th nearest neighborhood of the initially failing agent $i$).  It
nicely separates two effects that determine the severity of a cascade:
(a) the size of the initial shock $\phi_{\star}$, (b) the number of
neighbors to share the load and their respective safety margin,
i.e. $K(1-\alpha)$.  In the limit of large external shocks, and
independent of further assumptions about the
threshold distribution, the sequence of the critical thresholds
$\theta^{c}_{(t)}$ crucially depends on the sign of the factor
$K(1-\alpha)$. If $K(1-\alpha)>1$, the sequence $\theta^{c}_{(t)}$
will approach zero exponentially, i.e. with increasing distance from
the initially failing agent, this condition will be more easily
met. Hence, there should be a finite $t^{\prime}$ at which all
reasonably chosen threshold values $\theta_{r}$ are larger than the
critical threshold, which implies that the cascade stops. This is
shown in Fig.~\ref{fig:theta-c-n} for the case of the uniform
threshold distribution for different values of the safety margin
$(1-\alpha)$. We can verify for the given set of parameters that for
$\alpha=0.2$, and $\alpha=0.5$ the cascade stops right after $t=1$,
while for $\alpha=0.7$ it stops after $t=3$. On the other hand, for
$\alpha=0.8$ we see that the critical threshold is already at $t=3$
larger than any existing threshold, so the full cascade cannot be
prevented.

While this is an intuitive and illustrative example, we will calculate
analytically the exact time $t^{\prime}$ at which the cascade may
stop, in the following section. We note again that due to the finite
system size cascades may stop already at time $t^{\circ}$, which gives
an additional limit.

\section{Critical conditions for systemic risk}
\label{sec:risk}

\subsection{Analytical estimations of cascade sizes}
\label{sec:sizes}

Up to this point, we have derived a measure for systemic risk $X(t)$,
Eq.~\eqref{eq:xt} that is based on the fraction $f(t)$ of agents
failing at a given time $t$. Failure cascades can propagate in the
system if the net fragility of agents $\phi_{r}(t)-\theta_{r}$ is
positive, i.e. if the load exceeds the capacity. While $\phi_{r}(t)$
becomes a function of the redistribution of loads in previous time
steps, the capacity is determined by a threshold distribution function
$P(\theta)$, for which we use three different specifications. Already
the general framework outlined above allows us to expect ``infinite''
($X\to 1$) and ``finite'' ($X<1$) cascades, where the latter can
encompass the whole system or stop before. In the following, we will
specify the conditions for these findings for the different threshold
distributions.

\paragraph{Cascade size for homogeneous threshold}

Let us start with the simplest case that all agents have the same
threshold and the same number of neighbors. As stated above, there is
a failing agent $i$ at $t=0$, for which $\phi_i(0) = \phi_{\star}$.
Because of the homogeneous distribution, it can be noted that if an
agent $r$ fails due to the failure of one of its neighbors, $r^{\star}$,
then all the neighbors of $r^{\star}$ will fail as well. I.e. $f(t)=1$ if
$f(t-1)=1$. Let $\phi_{(t)}$ be the load of agents at a distance $t$
of the initially failing agent at $t=0$ and let us assume that agents
at a distance lower than $t$ already failed. Then, the load of agents
in shell $t$ is
\begin{equation}
\phi_{(t)} = \alpha \bar \theta + \frac{\phi_{(t-1)} }{K}.\label{eq:rec:homogeneous}
\end{equation}
With the initial condition $\phi_{(0)} = \phi_{\star}$, this recursive
equation can be easily solved, yielding
\begin{equation}
  \phi_{(t)} =   \alpha \bar\theta \, \frac{1-K^{-t}}{1-K^{-1}} + \frac{ \phi_{\star} }{K^t} .\label{eq:rec:homogeneous_solved}
\end{equation}
I.e., agents exposed to the redistribution of load at time $t$ will
fail if $\phi_{(t)}>\bar \theta$. This equation allows to gain insight
into the cascade mechanism. On the one hand, according to the above
discussion of the \textbf{EEE} regime, infinite cascades can only be triggered
if $K (1-\alpha) < 1$, irrespective of the threshold
distribution. This means a topological effect (the number and safety
margin of neighbors among which the load is redistributed) decides
about finite and infinite cascades.

On the other hand, when $K (1-\alpha) \geq 1$, the initial load
$\phi_{\star}$ cannot (by itself) trigger an infinite cascade in the
case of homogeneous threshold. I.e. even in the \textbf{EEE} regime, a cascade
will only last $t^{\star}$ time steps, where $t^{\star}$ results from the
failing condition $\theta^{(t^{\star})} \leq \phi_{(t^{\star})}$. From the condition $f(t^{\star})=0$, we can compute
\begin{equation}
t^{\star} = \frac{\log \left\{ (1-K^{-1}) \phi_{\star} - \alpha \bar\theta \right\} - \log \left\{ (1-K^{-1})\bar\theta  - \alpha \bar\theta \right\} }{\log K}.
\label{eq:tstar-uniform}
\end{equation}
As discussed before, the actual time $t^{\prime}$ at which the cascade
stops is $t^{\prime} = \min(t^\circ,t^{\star})$, where $t^\circ$ denotes the
time where the cascade reaches the system size, Eq. (\ref{tcirc}).

Knowing $t^{\prime}$, we can further calculate the {\em systemic risk}
according to Eq. (\ref{eq:xt}), with $f(t)=1$, i.e. $F(t) = K(t)$ for $t\leq t^{\prime}$ and
$F(t) = 0$, otherwise. We find
\begin{equation}
X(t^{\prime})=\frac{K^{t^{\prime}-1}-1}{N(K-1)}
\label{eq:xhom-bethe}
\end{equation}
for a Bethe lattice or a tree with coordination number $K$ and
\begin{equation}
X(t^{\prime}) \propto \frac{K (t^{\prime}+1) t^{\prime}}{N}
\label{eq:xhom-reg}
\end{equation}
for a regular lattice, with the exact factor depending on the
topology.

\paragraph{Cascade size for uniform threshold distribution}

We now turn to the simplest case that allows some heterogeneity in the
agent's threshold, which is the uniform distribution
$P(\theta)=U(\bar \theta-\sigma,\bar \theta+\sigma)$. The failing condition
in Eq.~(\ref{theta_c}) for the nearest neighbors still holds, but the
question is how often we find thresholds below the critical limit:
\begin{equation}
  f(1)=\int_{\bar \theta-\sigma}^{\theta_{(1)}^{c}} P(\theta) d\theta
= \frac{\theta_{(1)}^{c}-(\bar \theta-\sigma)}{2\sigma}.\label{f_1_null}
\end{equation}
With $\theta^c_{(1)}$ given by Eq. (\ref{theta_c}) it turns out that
the fraction of failing agents at the first time step is
\begin{equation}
  f(1)= \frac{\phi_{\star}/[K(1-\alpha)]-(\bar \theta-\sigma)}{2\sigma}
  \quad \mathrm{if} \quad \phi_{\star}/(\bar \theta - \sigma)  \geq K(1-\alpha)
  \label{eq:f1_uniform}
\end{equation}
and  $f(1)=0$ otherwise.
Regarding $f(t)$, we know from Eq.~(\ref{f_t_general}) that the
fraction of failing agents at any time step crucially depends on the
history of failed agents, i.e.~the path connecting the initially
failed agent with the currently failing one.  Therefore, in general,
the process is not solvable.  There is the need of a simplifying
assumption to break the integral expression in Eq.~\eqref{f_t_general}
into solvable parts.

For the \textbf{EEE} regime we use Eq.~(\ref{eq:theta_c_t-EEE}) and
the underlying assumptions to obtain the closed equation,
\begin{equation}
  f(t)= \int_{\bar \theta - \sigma}^{\theta_c(\phi_{(t-1)})} d\theta P(\theta)
   =\frac{1}{[K(1-\alpha)]^t } \frac{\phi_{\star}}{2\sigma}-\frac{\bar \theta-\sigma}{2\sigma}.
\label{eq:fn}
\end{equation}
With this expression, we find from $f(t^{\star})=0$ the time at which the cascade stops   for the uniform threshold distribution:
\begin{equation}
  t^{\star}=\frac{\log (\phi_{\star})-\log(\bar \theta-\sigma)} {\log [K(1-\alpha)]}.
\label{tprimeuniform}
\end{equation}
Again, the  time at which the cascade ends is given by $t^{\prime}=\min(t^\circ,t^{\star})$, with $t^{\circ}$ given by Eq. (\ref{tcirc}).

Considering instead the \textbf{RIE} regime, $\phi^{\star}\sim
\bar{\theta}$, where redistribution effects play a mayor role, we use
Eq. (\ref{eq:rec:aux}) and the underlying assumptions. Neglecting
capacity-capacity correlations among agents, we find for the case of
the uniform threshold distribution:
\begin{equation}
  f(t) = \frac{ \left( \theta^c_{(t-1)} - \bar\theta + \sigma \right)  \left[ \theta^c_{(t-1)} +  \left[  1 - 2 K (1-\alpha) \right] ( \bar\theta - \sigma )\right]   }{8\sigma^2\ K(1-\alpha)}. \label{eq:rec:red2}
\end{equation}
 The time at which the cascade stops is, as in the previous cases, given by the condition $t'=\min(t^\circ, t^\star)$, where $t^\star$ is the first time step that verifies  $f(t^{\star})=0$, and $t^\circ$ is the one defined in Eq.~\eqref{tcirc}.

\paragraph{Cascade size for power-law threshold distribution}

Now, we discuss the case where agent's threshold follows a power-law
distribution, Eq. (\ref{eq:power}), and can vary by orders of
magnitude. With the same procedure as used before, we determine the
fraction of failed agents during the initial cascade as
\begin{equation}
  f(1)=\int_{\theta_{\mathrm{min}}}^{\theta_{(1)}^{c}} P(\theta) d\theta
  = 1-\left( \frac{\phi_{\star}/\theta_{\mathrm{min}}}{K(1-\alpha)}\right)^{1-\gamma}. \label{f_1_power}
\end{equation}
So, cascades are obtained if $\phi_{\star}/\theta_{\mathrm{min}} \geq K(1-\alpha)$.

Considering first the \textbf{EEE} regime, we use the approximation given by  Eq.~\eqref{eq:theta-c-n}
and find for the fraction of failing agents during time step $t$:
\begin{equation}
f(t)=1-\left( \frac{\phi_{\star}}{\theta_{\min}} \right)^{1-\gamma} \left[ K(1-\alpha) \right]^{(\gamma-1)t}.
  \label{eq:fn_power}
\end{equation}
From $f(t^{\star})=0$, we calculate the time when the cascade stops as
\begin{equation}
  \label{eq:tprimepower}
 t^{\star}=\frac{\log (\phi_{\star}) - \log (\theta_{\mathrm{min}})}{\log [K(1-\alpha)]}.
 \end{equation}
Again, following our previous discussion, the cascade stops at $t^{\prime}=\min(t^{\star},t^\circ)$.

In the \textbf{RIE} regime, on the other hand, Eq.
, (\ref{eq:rec:aux}) has to be applied to the
power law distribution, to yield
\begin{equation}
 f(t) =   1 - \left( \frac{ \theta_{\min} }{ \theta^c_{(t-1)} } \right)^{\gamma-1}  - \frac{[K(1-\alpha)^{\gamma-1}]}{2} \left[ 1 - \left( \frac{ \theta_{\min} }{ \theta^c_{(t-1)}}\right)^{2\gamma-2}\right].
\end{equation}
which, together with the failure condition of Eq. (\ref{eq:rec:red1})
provides a set of recursive equations, to be solved numerically.

\subsection{Numerical results for the EEE regime}
\label{sec:het}

\paragraph{Size of finite cascades}

The analytical results for $f(t^{\star})$ and
$t^{\prime}=\mathrm{min}(t^{\star},t^{\circ})$ allow us now to
calculate the systemic risk $X(t^{\prime})$ for the different
threshold distributions, by varying system parameters such as the
safety margin $(1-\alpha)$ or the network topology. In this section,
we first concentrate on the \textbf{EEE} regime, where the external
shock is comparable to the network capacity and much larger than the
average threshold of an agent, $\phi^{\star}\sim Q \geq \bar{\theta}$.

\begin{figure}
\centering
\includegraphics[width=0.45\textwidth,angle=0]{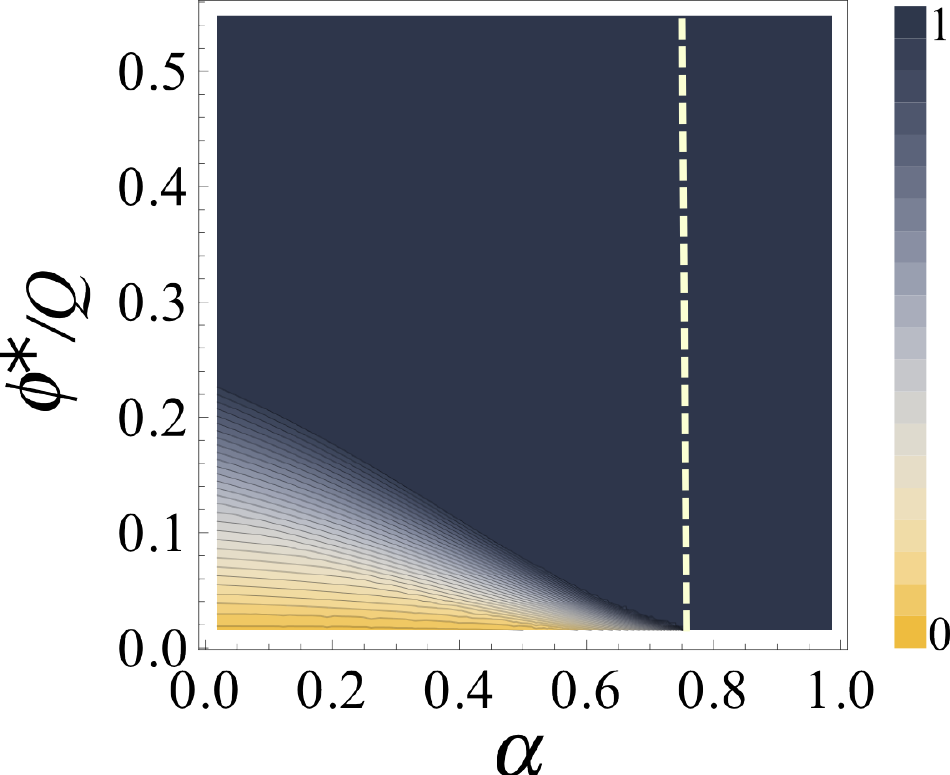}
\includegraphics[width=0.45\textwidth,angle=0]{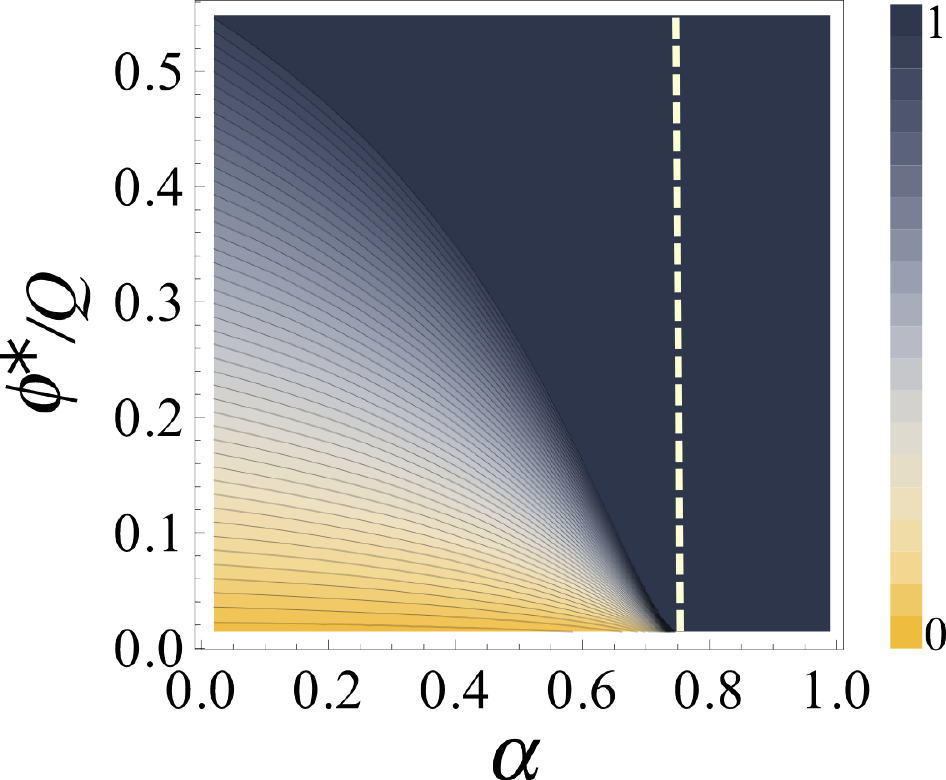}
\caption{
Systemic risk in a system with uniform threshold distribution
  with $\bar{\theta}=1$, and $\sigma=0.5.$
The contour plots show $X$
  dependent on the initial shock $\phi_{\star}$ normalized to the
  network capacity $Q$ for different values of $\alpha$. Left: Bethe
  lattice, right: regular network.  $K=4$, so the dashed line
  $K(1-\alpha)<1$ separates the region of ``infinite'' cascades,
  $X(t^{\prime})=1$, from the region of possibly finite cascades,
  $X(t^{\prime})<1$.
 \label{fig:X-uniform}
}
\end{figure}
In Fig.~\ref{fig:X-uniform} we compare, for the uniform threshold
distribution, the systemic risk $X$ in Bethe and 2D regular
lattices. We remind that the difference is in the number of agents
potentially affected by the cascade at a given time $t$. For Bethe
lattices and regular trees, we have $K(t)= K^{t}$, whereas for regular
networks $K(t)\propto K t$, i.e. for a given $t$ in Bethe lattices
much more agents are affected. Conversely, for a given $N$, regular
lattices have a larger diameter. For example for the 2D regular
lattice the diameter grows with system size as $\sqrt N$. On the other
hand, the diameter in a Bethe lattice grows as $\log N$.

In both cases, trivially, if the safety margin $(1-\alpha)$ vanishes,
any external shock results in an immediate collapse. This also happens
for finite safety margins as long as the global stability condition
$K(1-\alpha)\geq 1$ is not met. On the other hand, for large safety
margins, $\alpha \to 0$, we do expect finite cascades. As the plots
show, the parameter region for these is much larger for regular
networks where at each time step a smaller number of agents is
affected, than for Bethe lattices. As shown, in the first case an
initial shock of almost 30\% the network capacity already leads to
full cascade, whereas in the second cease it requires an initial shock
of almost 60\% the system's available capacity for a full cascade.

\begin{figure}
\centering
\includegraphics[width=0.45\textwidth,angle=0]{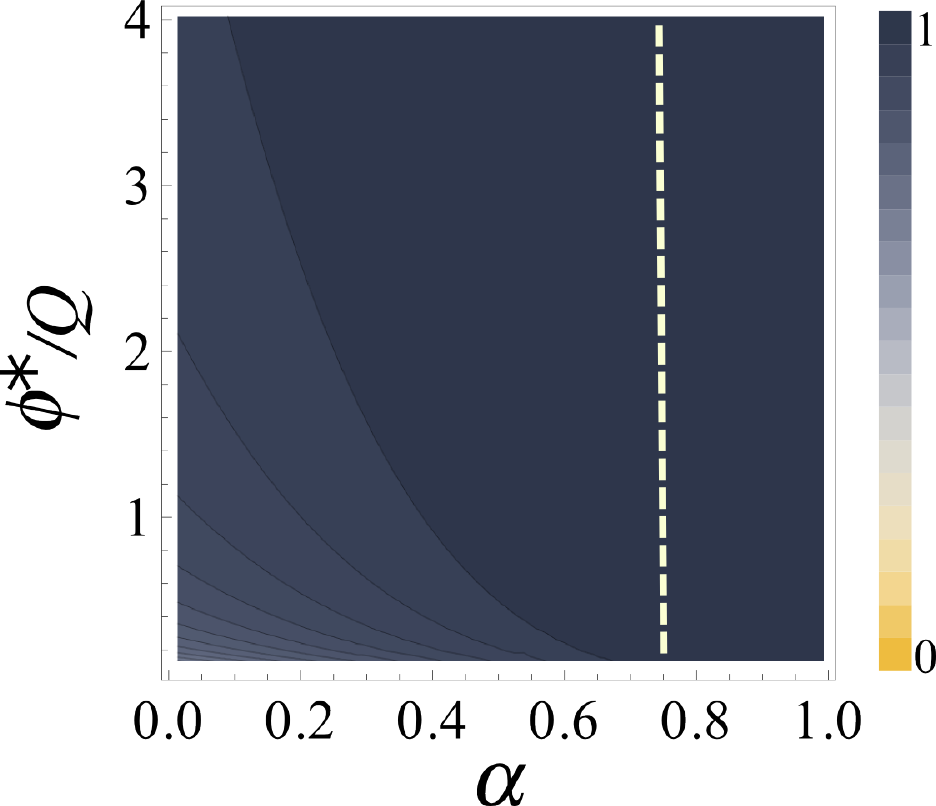}
\includegraphics[width=0.45\textwidth,angle=0]{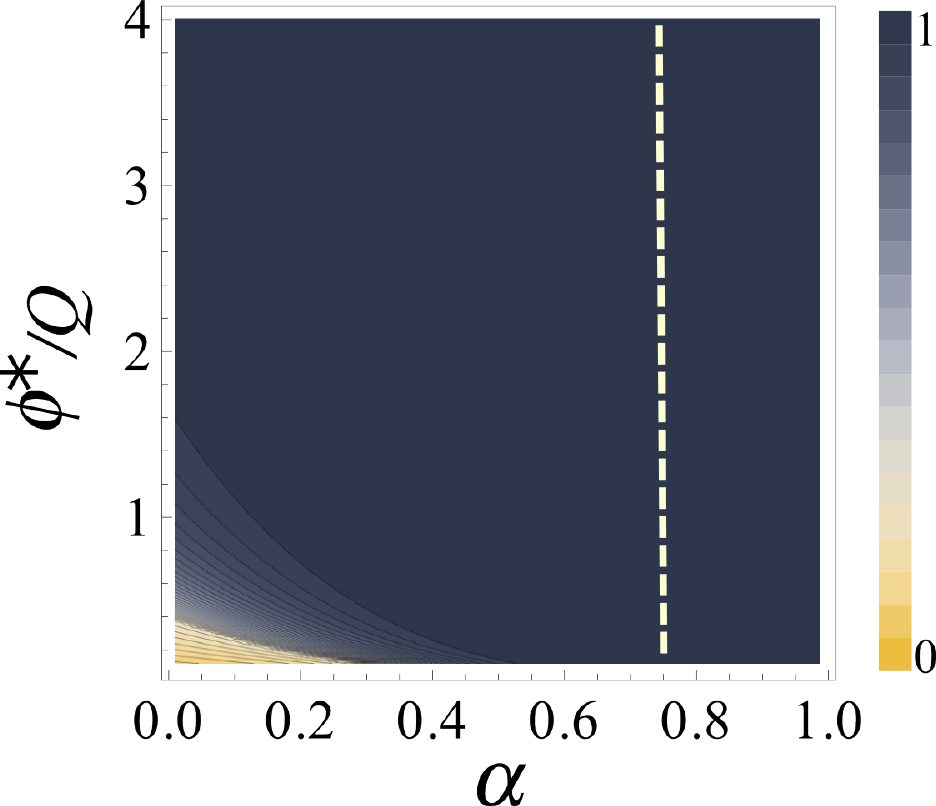}
\caption{Systemic risk in a system with power law threshold distribution: (left) $\gamma=1.5$, (right) $\gamma=3.0$, for a Bethe lattice with $K=4$, $N=1000$. Cf. also Fig. \ref{fig:X-uniform}.
}
\label{fig:X-plaw}
\end{figure}
The results are to be compared with Fig.~\ref{fig:X-plaw}, where we
plot the systemic risk $X$, for Bethe lattices only, for a power law
threshold distributions with two different exponents $\gamma$.  We
remind that the network capacity $Q$ is comparable to the case of the
uniform threshold distribution only for $\gamma=3$,
Fig.~\ref{fig:X-plaw} (right) and we also observe a similar dependence
of $X$ on the safety margin $(1-\alpha)$ and on the relative initial
load $\phi_{\star}/Q$. To be precise, in this case the safety margin
plays a less important role, but we find finite cascades also for
$\phi_{\star}/Q>0.5$, which was not the case for the uniform threshold
distribution.

The situation differs for $\gamma=1.5$, as Fig.~\ref{fig:X-plaw}
(left) shows. Here the system seems to be much more vulnerable,
indicated by large values of $X$ in all parameter regions. To put this
finding into perspective, we remind that for $\gamma<2$ the network
capacity is much larger than for $\gamma>2$, i.e. the initial shock
also has much higher values as compared to Fig.~\ref{fig:X-plaw}
(right). This explains the severity of the cascades in this case.

The results obtained for these two particular values of $\gamma$ can
be generalized to $\gamma\leq 2$ and $\gamma>2$. As a consequence, we
may conclude that with a much broader threshold distribution, the
system can absorb higher initial shocks (in absolute values), but
shocks of a size comparable to the network capacity most likely result
in infinite cascades, i.e. total failure.

\begin{figure}
\centering
\includegraphics[width=0.42\textwidth,angle=0]{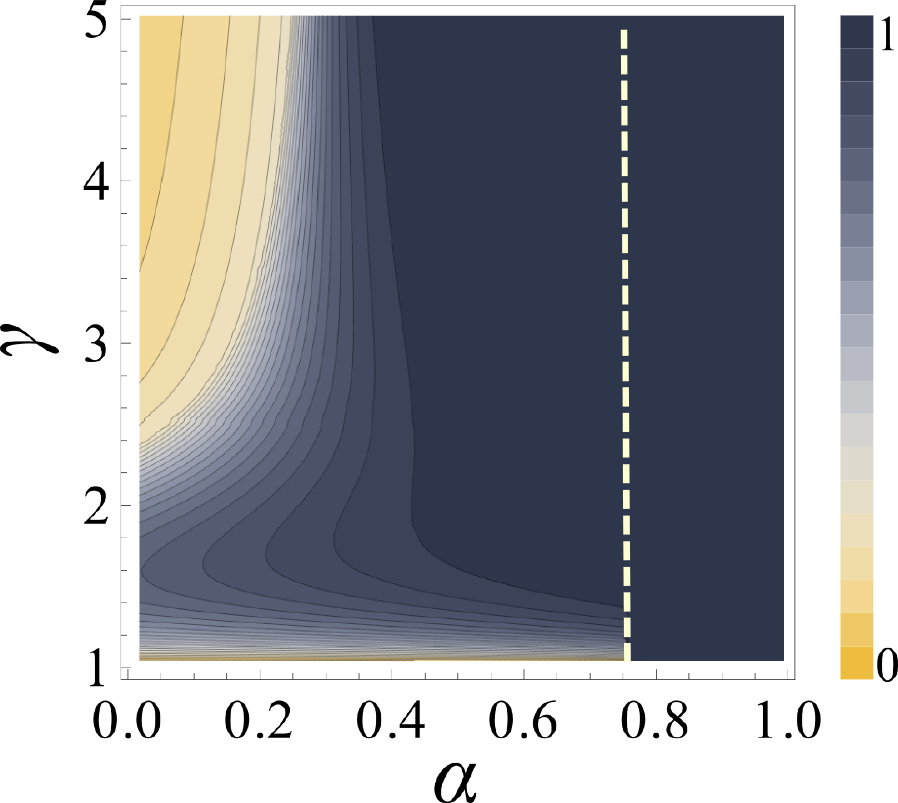}
\includegraphics[width=0.42\textwidth,angle=0]{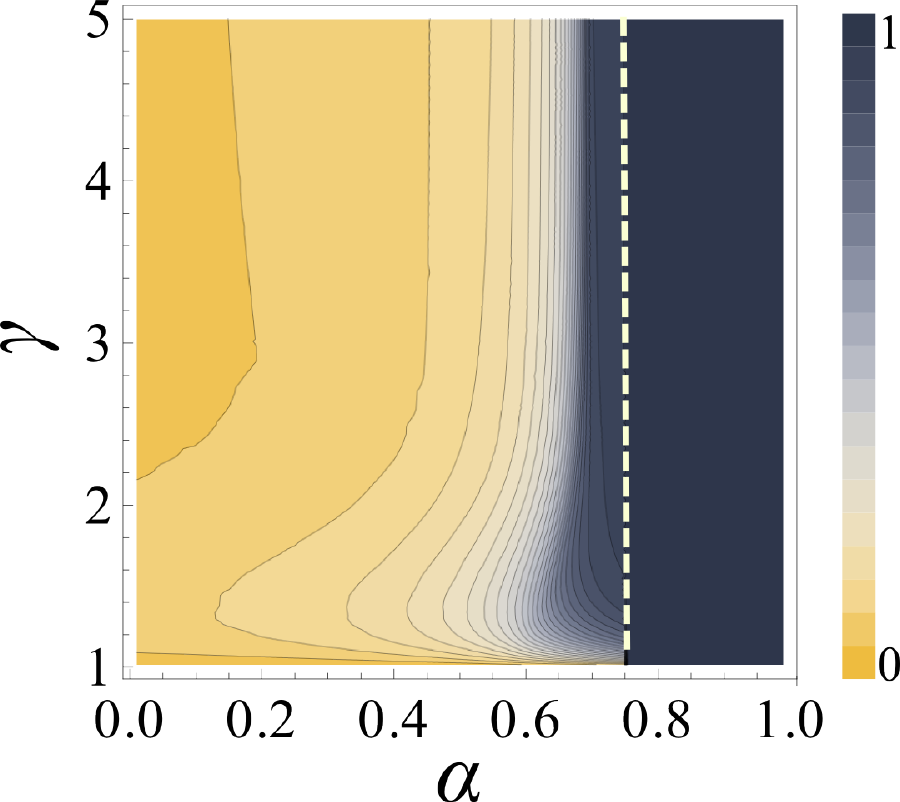}
\caption{Systemic risk in a system with power law threshold
  distribution. The relative initial shock was fixed:
  $\phi_{\star}/Q=0.2$. $N=1000$, $K=4$. (Left) Bethe lattice, (right)
  2D regular lattice. Cf.~also Fig. \ref{fig:X-plaw}.}
\label{fig:X-gammas}
\end{figure}
To better understand the role of the skewness of the threshold
distribution and the topology, we fix the relative initial shock
$\phi_{\star}/Q=0.2$ and vary $\gamma$ for two different network
topologies.  Fig.~\ref{fig:X-gammas} confirms (a) that a Bethe lattice
or regular tree structure leads to more severe cascades as compared to
a regular network, which is due to the smaller diameter of the
network, and (b) that an increasing skewness, i.e. smaller values of
$\gamma$ lead to an increasing systemic risk. Remarkably, there is a
non-monotonic dependence of $X$ on $\alpha$ and $\gamma$, and the
cascade size becomes larger around $\gamma\approx1.5$.  The reason for
this is, on the one hand, the system size dependence of the network
capacity for $\gamma\leq 2$ and, on the other hand, the larger
fragility resulting from a more skewed distribution (i.e. thresholds
are closer to $\theta_{\mathrm{min}}$). The first effect is a global
one, i.e. larger load is added to the system, the second is a local
one, i.e. there are fewer agents that can handle large loads.

\subsection{Results for the RIE regime
}\label{sec:sysnodes}

The previous results have focused on the \textbf{EEE} regime of
large external shocks, $\phi_{\star}\sim Q \gg \bar{\theta}$. This means
that the initial load is largely responsible for triggering the
cascades. Here we focus on the opposite case, the \textbf{RIE} regime,
$Q \gg \phi_{\star} \sim \bar \theta$, where small
fluctuations of an agent's load  lead to the agent's failure provided that
the safety margin of that agent was rather small. The question is then
under which conditions this failure leads to a cascade of macroscopic
size.  Applications of this case include power-grid cascade failures
\cite{motter2002cascade,crucitti2004model,CRUCITTI_cascades_power_grid}, or failures of server infrastructure
\cite{rossi2009}.

We now consider $\phi_{\star} = \theta_i$ for the failing agent,
and we assume that the system is in a critical condition. This means that a single failure   among the neighbors of the initially failing agent (i.e.~$f(1) \geq 1/K$) is enough to trigger a system-wide cascade.
We now define $\phi_\star^{c}$ as the load such that $f(1)$ is exactly equal to $ 1/K$.
With these assumptions, we compute the ensemble average $\mean{X}$ of
the systemic risk
\begin{equation}
\label{eq:F_inf_arbitrary}
  \langle X \rangle =\int_{\phi_\star^{c} } d\theta_{i} \, P(\theta_i). \end{equation}
The integral runs over the threshold $\theta_{i}\geq \phi_\star^{c}$
of all possible agents $i$ which trigger a full cascade.  These agents
can be regarded as systemically important because their failure
induces a systemic collapse.

The quantity $\langle X\rangle$ then represents the frequency at which
a randomly chosen agent can trigger the complete failure of the
system. In the following we compute $\langle X \rangle$ for the
uniform and power-law thresholds distribution.

\paragraph{Uniform threshold distribution}
The \textit{initial critical load} $\phi_{\star}^{c}$ in the
\textbf{RIE}  regime can be easily computed from
Eq.~\eqref{eq:f1_uniform}, using the condition $f(1) = 1/K$:
\begin{eqnarray}
\label{eq:uniform_critical_load_RIE}
\phi_{\star}^{c}= K(1-\alpha) \left[ {2 \sigma}/{K}+(\bar \theta-\sigma) \right].
\end{eqnarray}
Then, the  average systemic risk for the uniform distribution $\mean{X_{u}}$ is given by
\begin{eqnarray}
\label{eq:X_uniform}
\mean{X_{u}} = \begin{cases}
 0 & \mathrm{if} \quad \bar \theta - \sigma >\phi^c_{\star}  \\
              \left[ (2\alpha-1)  - K(1-\alpha) \right] / 2 - \bar \theta \left[ K (1-\alpha) -1 \right] /{2\sigma} & \mathrm{if} \quad  \bar \theta - \sigma \leq \phi^c_{\star} \leq \bar \theta +\sigma \\
1 & \mathrm{if} \quad  \phi^c_{\star} > \bar \theta +\sigma
               \end{cases}.
\end{eqnarray}

\paragraph{Power-law threshold distribution }
In an analogous way we obtain from Eq. (\ref{f_1_power}) with $f(1)=1/K$ the
expression for the \textit{initial critical load} $\phi_{\star}^{c}$ in the case of a
power-law threshold distribution:
\begin{equation}
\label{eq:power_law_critical_load_RIE}
\phi_{\star}^{c} = \theta_{\min}K(1-\alpha)(1-K^{-1})^{-{1}/{\gamma-1}}.
\end{equation}
This allows us to calculate the average systemic risk as:
\begin{equation}
\label{eq:X_power_law}
\langle X_{p} \rangle = \begin{cases}
\frac{1-{K}^{-1}}{\left[ K (1-\alpha)\right]^{\gamma-1}} & \text{if} \quad \phi^c_{\star} \geq \theta_{\min} \\
1 & \text{otherwise}
  \end{cases}.
\end{equation}

\begin{figure}
  \centering
  \includegraphics[width=0.45\textwidth,angle=0]{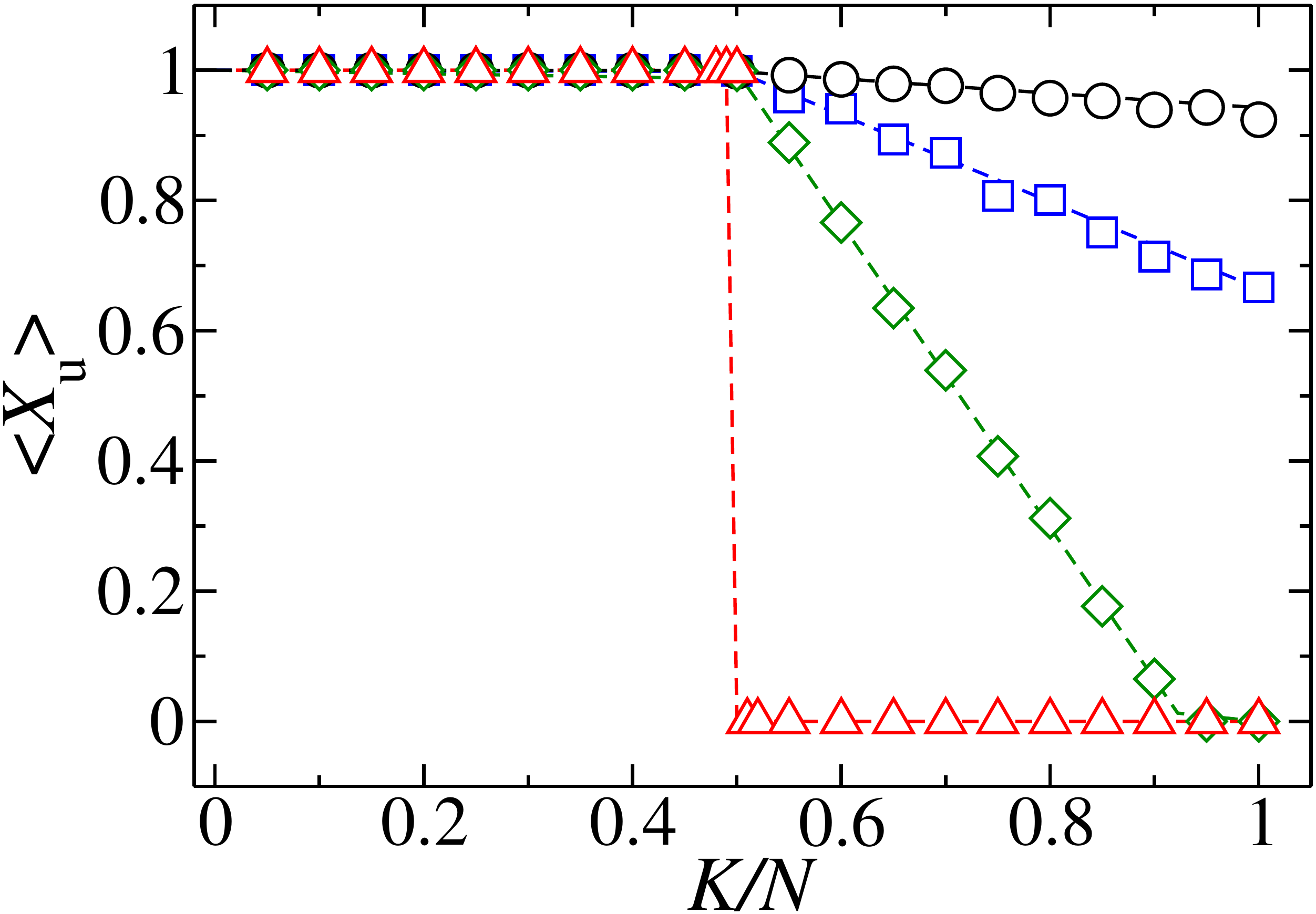}
  \includegraphics[width=0.45\textwidth,angle=0]{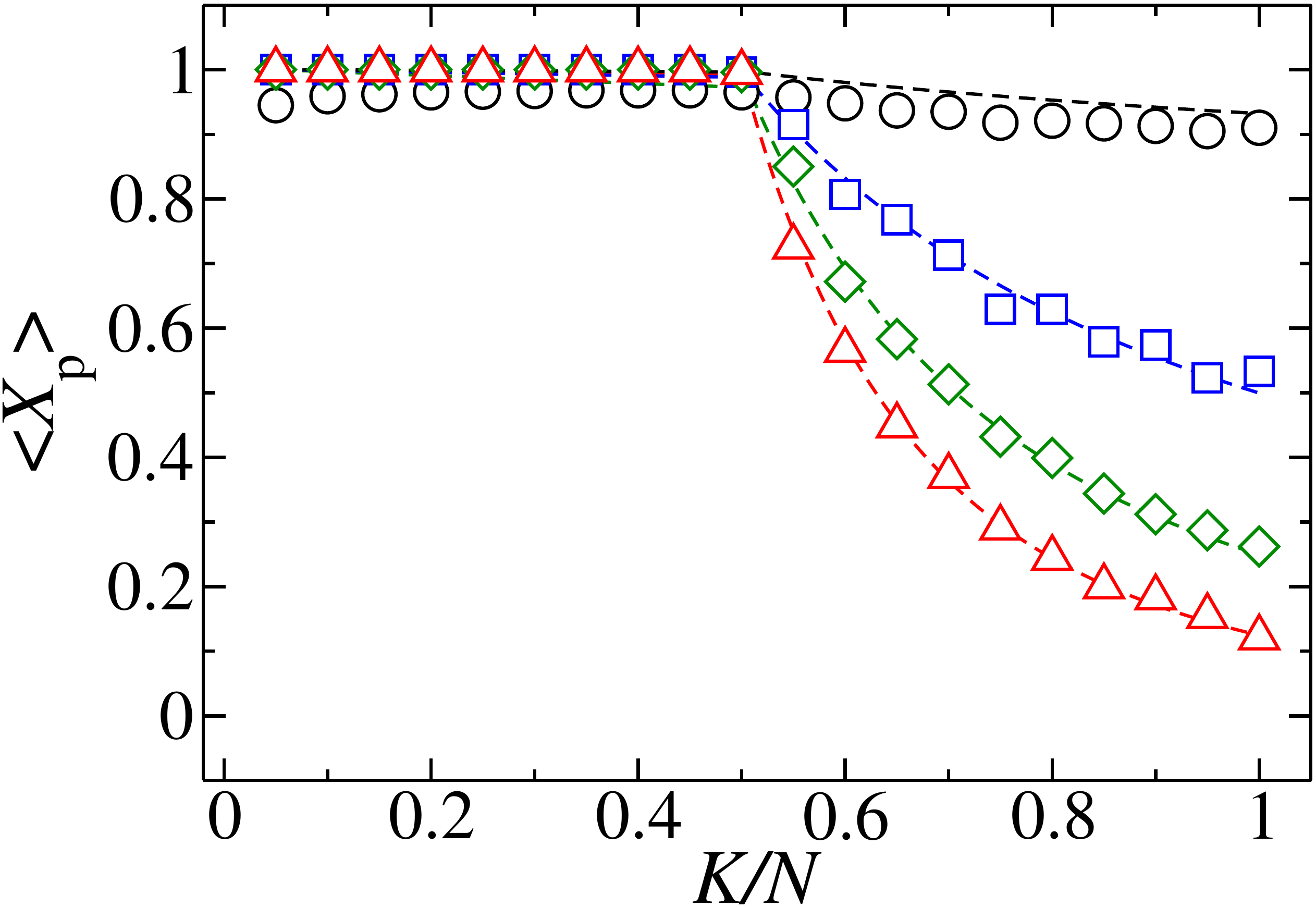}
  \caption{
Frequency of ``infinite'' cascades $\mean{X}$ for the
  uniform (left) and the power law (right) threshold distribution in the \textbf{RIE} regime.
  We consider regular random networks of size $N=1000$ with varying $K$; the
  safety margin of agents is  set to $(1-\alpha) = 0.002$. Open symbols
  show  simulation results,  dashed lines  analytic results. Parameters
for the uniform distribution: $\bar{\theta}=1$,  $\sigma=0$ (triangles), $\sigma=0.3$
  (diamonds), $\sigma=0.6$ (squares), and $\sigma=0.9$ (circles). Parameters for
  the power law distribution:
  $\gamma=1.1$ (circles), $\gamma=2$ (squares), $\gamma=3$ (diamonds),
  and $\gamma=4$ (triangles).
}
\label{fig:Cascade_regime}
\end{figure}

In order to study the role of connectivity in the cascade process, we
created \emph{random regular networks} \cite{construction} with
arbitrary values of the average connectivity $K/N$.
Fig.~\ref{fig:Cascade_regime} shows the average cascade size $\langle
X \rangle$ for a rather small safety margin, which is in line with the
\textbf{RIE}  regime. This means that small fluctuations of the load of
a single agent may lead to its failure. Fig.~\ref{fig:Cascade_regime}
allows to compare the analytical expressions in
Eqs.~\eqref{eq:X_uniform}, \eqref{eq:X_power_law} with numerical
simulations of the cascade process.
The graphs show a sharp transitions from $\mean{X}=1$ to $\mean{X}<1$
at $K/N=1/2$. This results directly from the change in global
instability $K(1-\alpha) \leq 1 $, at that particular point.
On the other hand, when the system is not globally unstable, the
results show that only a subset of agents are able to trigger a
cascade in the system, their fraction indicated by $\mean{X}$

The left panel of Fig.~\ref{fig:Cascade_regime} shows the results for
the uniform threshold distribution.  For identical agents (left panel,
$\sigma=0$), a sharp transition between complete failure
($\mean{X}=1$) and no failure ($\mean{X}=0$) can be observed. This
result immediately follows from Eq.~\eqref{eq:rec:homogeneous_solved},
in the limiting case $t \to \infty$.  For a fixed value $K/N > 1/2$,
the graphs show that larger values of $\sigma$ exhibit larger
frequencies of full cascades, $\langle X\rangle$. This results from
Eq.~\eqref{eq:uniform_critical_load_RIE} which shows that the critical
initial load $\phi^c_{\star}$ decreases for increasing heterogeneity;
thus, for larger values of $\sigma$ a larger fraction of agents is
likely above such a threshold.

The right panel of Fig.~\ref{fig:Cascade_regime} shows the results
for the power-law threshold distribution.  Again, broader
distributions, i.e.~lower values of $\gamma$, result in a higher
probability of complete failures. This is in line with
Eq.~\eqref{eq:power_law_critical_load_RIE} where, for a fixed $K$, the
critical load increases with $\gamma$. At the same time, the threshold
distributions become more narrow with larger $\gamma$. Thus, the
amount of {\em systemically important} agents that are able to trigger
a full cascade, is much lower in distributions with large $\gamma$,
and thus the average systemic risk decreases.

\section{Conclusions} \label{sec:concl}

The model proposed in this paper is based on very simple ingredients,
to allow for analytical treatment: (i) a regular network, i.e. all
agents have the same number of neighbors, $K$, (ii) a constant safety
margin $(1-\alpha)$, the same for all agents, which defines a fixed
relation between the load $\phi_{r}$ an agent can possibly carry and
the threshold $\theta_{r}$ at which it fails, (iii) an initial
condition that only one randomly chosen agent $i$ fails when facing a
load $\phi_{\star}$, (iv) a redistribution of the load of the failing
agent to its $K$ neighbors.

The variability of the model comes from two assumptions: (v) the
threshold distribution $P(\theta)$ which was chosen as a delta
distribution, a uniform distribution, or a power law distribution,
(vi) the severity of the initial shock, which was either of the order
of the network capacity $Q$, i.e. much larger than the average
threshold, or much smaller than the network capacity, i.e. of the
order of the average threshold. The latter allowed us to distinguish
between two important regimes: (a) \textbf{EEE} regime, where an
extreme external event was large enough to cause the failure of many
agents, (b) \textbf{RIE} regime, where small random internal events
may result in the failure of a single agent. In both cases, this
initial failure may have triggered a cascade of failures in the
neighboring agents.

For both regimes, we are interested in the following question: what is
the possible size of a failure cascade, $X(t)$, measured as the total
number of failed agents compared to the system size $N$, at a given
time $t$. Will this be an ``infinite'' ($X\to 1$) or a ``finite''
($X<1$) cascade and, in the latter case, at what time will it stop: at
a time $t^{\circ}$ where the cascade has already reached all agents
but did not cause all of them to fail, or at a time $t^{\star}$ before
it has passed the entire system. We regard $X$ as a measure of
systemic risk.

For both regimes, we are able to derive analytical results to answer
these questions. These results allow us to draw conclusions about the
conditions that lead a smaller systemic risk. In the following we summarize our finding:

(1) We derive a global stability condition $K(1-\alpha)\geq 1$ that
has to be met in order to allow for finite cascades, in principle. The
larger the number of neighboring agents, $K$, or the larger the safety
margin, $(1-\alpha)$, the more likely this condition is met. This
allows for an interesting discussion because of the possible trade-off
between the two ingredients. In most cases, the safety margin is given
by the \emph{technical constitution} of an agent, e.g. in power grids
or routing servers. $K$, on the other hand, refers to the
\emph{network topology} but not to internal properties of
agents. Hence, systemic risk can be reduced by increasing the network
density - at least up to a certain point \cite{Battiston2009a}. It should
be noted that we have assumed the initial failure of only one agent,
here. If it is, however, more costly to improve the network
connectivity than increasing the safety margin of the agents, the
latter can serve the same purpose, namely reducing systemic risk.

(2) In a system of thousands of agents, the network capacity, $Q$,
which is the total load the system can carry a priori, is also quite
big. One would not easily assume that an initial shock $\phi_{\star}$
is of the same magnitude as $Q$ as in the \textbf{EEE} regime. Hence
such big shocks are extreme external events to the system. The
interesting finding in this paper is that even such extreme events may
not lead to an ``infinite'' cascade, i.e. to a total collapse of the
system. Instead, provided that the global stability condition is met,
we find a broad range of system parameters where such cascades stop at
finite time, affecting only part of the system. We have shown that
systemic risk resulting from extreme shocks can be reduced (a) by a
regular lattice structure (as opposed to e.g. a regular tree
structure), (b) by a broad threshold distribution. In the latter case,
we found finite cascades, i.e. $X<1$, even if the initial shock was
\emph{2-4 times larger} than the total network capacity, which can be
regarded as a sign of real robustness. Comparing the power-law
threshold distributions of $\gamma=1.5$ and $\gamma=3$, we found that,
in absolute measures of the shock, the broader distribution lead to
the more robust systems. In relative measures, however, this result
inverts, simply because a broader distribution also results for a
larger network capacity, and hence for larger initial shocks, while
the relative measure remains the same.

(3) Investigating the \textbf{RIE} regime where the initial shock was
of the order of the threshold of an average agent, i.e. much smaller
than the total network capacity, a systemic failure can occur only if
(a) the the initial shock is larger or equal to the threshold of the
initially failing agent, and (b) the redistribution of load is large
enough. Hence, dependent on the threshold distribution we can
calculate this failure probability. Even in the global stability
regime, we find ``infinite'' cascades, but the probability of their
occurrence depends on the probability that randomly chosen agent fails
initially. The broader the threshold distribution, the more likely
this condition is met, i.e. the frequency of observing ``infinite''
cascades increases with the heterogeneity.

(4) The initial question: ``How big is too big'', from this
perspective, can be answered as follows: Initial shocks, even if they
exceed the capacity of the whole system (not just the capacity of a
single agent), are probably not the problem. Of course, there are
parameter regimes that lead to complete collapse ($X\to 1$). At the
same time, we see that a change of $\phi_{\star}$ of 10 or even 50
percent does not change the systemic risk very much. Of much larger
influence are system parameters related to the network topology, the
safety margin, and the threshold distribution. As it was also found in
other papers \cite{Lorenz2009b}, an optimal heterogeneity in the
agent's threshold can reduce systemic risk considerably. In addition
to that we find that an change of the safety margin by 10 or 50
percent generates a much larger impact on systemic risk than a
comparable change in the external shock. So, when seeking for
protection against systemic risk the focus should be (a) on those
parameters that influence the global stability, i.e. $K(1-\alpha)$
(see above), and (b) on the optimal heterogeneity in the threshold
distribution.

\end{document}